A New Multiferroic State with Large Electric Polarization in Tensile Strained $TbMnO_3$

Y. S. Hou, J. H. Yang, X. G. Gong[*], and H. J. Xiang[*]

[1] Key Laboratory of Computational Physical Sciences (Ministry of Education), State Key Laboratory of Surface Physics, and Department of Physics, Fudan University, Shanghai 200433, P. R. China

Abstract

By performing first-principles calculations, we systematically explore the effect of epitaxial strain on the structure and properties of multiferroic $TbMnO_3$. We show that, although the unstrained bulk $TbMnO_3$ displays a non-collinear antiferromagnetic spin order, $TbMnO_3$ can be ferromagnetic under compressive strain, in agreement with the experimental results on $TbMnO_3$ grown on $SrTiO_3$. By increasing the tensile strain up to 5%, we predict that $TbMnO_3$ transforms into a new multiferroic state with a large ferroelectric polarization，two orders of magnitude larger than that in the unstrained bulk, and with a relatively high Neel temperature E-type antiferromagnetic order. We also find that the ferroelectric domain and antiferromagnetic domain are interlocked with each other, thus an external electric field can switch the ferroelectric domain and the antiferromagnetic domain simultaneously. Our work demonstrates that strain engineering can be used to improve the multiferroic properties of $TbMnO_3$.

Multiferroic materials [1-2], displaying two or more orders of magnetic, polar, and elastic order simultaneously, have attracted much attention due to their potential applications as novel devices [3]. In these materials, the spin order and electric polarization can be controlled by the external electric field and magnetic field, respectively [4]. TbMnO$_3$, as a typical orthorhombic perovskite multiferroic material, has been the research topic of many studies, since Kimura *et al.* discovered the switch of ferroelectric polarization by magnetic fields [5]. The spins of Mn$^{3+}$ ions in bulk *Pbnm* TbMnO$_3$ form an incommensurate cycloidal spin spiral below $T_{lock} \approx 28\ K$ [6], and a polarization of $\sim 0.06\ \mu C/cm^2$ simultaneously appears along *c*-direction [7]. The polarization was found to be a consequence of spin-orbit coupling [8-9] and dominated by the contributions from the ion displacements [10-12]. Although there is a strong intrinsic magnetoelectric coupling in bulk TbMnO$_3$, the transition temperature is too low and the electric polarization is too small for realistic applications.

Recently, it has been demonstrated experimentally that unexpected anomalous ferromagnetism appears in (001)-oriented TbMnO$_3$ thin film epitaxially grown on the (001) plane of the cubic substrate SrTiO$_3$ (STO) [13]. Besides, RMnO$_3$ (R represents a rare-earth element) displays various spin orders with decreasing ion radius of R, i.e., LaMnO$_3$, TbMnO$_3$, and HoMnO$_3$ display the A-type antiferromagnetic (AFM) order, the non-collinear spiral order, and the E-type AFM order, respectively [14-15]. In addition, it was found that epitaxial strain could be used to tune the magnetic and polar order of perovskite magnetic oxides such as EuTiO$_3$ [16-17].

In this Letter, we propose to use epitaxial strain to tune the multiferroic properties of TbMnO$_3$. Our density functional calculations show that when TbMnO$_3$ is grown on SrTiO$_3$, the ground state is ferromagnetic (FM) with the centrosymmetric orthorhombic *Pbnm* space group, in agreement with experiments [13]. When the epitaxial tensile strain reaches 5%, the ground state of TbMnO$_3$ transforms into a new multiferroic state, i.e., insulating E-type AFM state with polar orthorhombic *Pmc2$_1$* space group. This new *Pmc2$_1$*-E-AFM state is characterized by a large electric polarization of 4.56 μC/cm$^2$ and a relatively high Neel transition temperature above 100 *K*. The E-AFM spin order is interlocked with the polar order, thus we propose a feasible strategy to switch the AFM domains through an applied electric field.

Our study is based on first-principles density functional theory (DFT) calculations [18]. We assume that TbMnO$_3$ is grown on the (001)-oriented cubic substrates. Hence epitaxial strain is defined as $\eta = \frac{a-a_0}{a_0}$ where *a* and $a_0$ are the in-plane lattice constants of the epitaxially strained and free-standing TbMnO$_3$ in cubic state [19]. As usual, we consider several common magnetic orders in perovskite, namely FM, A-type AFM (A-AFM), C-type AFM (C-AFM), G-type AFM (G-AFM), and E-type AFM (E-AFM) spin orderings. We double the unit cell in *a*- or *b*-direction to allow E-AFM magnetic order. For each strain and each magnetic order, structural relaxation begins with the initial structure with *Pbnm* space group and the in-plane lattice constants are fixed while the out-of-plane lattice constant and internal ionic coordinates are fully relaxed without imposing any symmetry. This strategy was shown to be successful in finding the ground state of the epitaxially strained perovskite oxides such as $SrRuO_3$ and $BiFeO_3$ [20]. For the E-AFM order, the unit cell is energetically preferred to be doubled along the *b*-axis in the case of compressive and small tensile strain, while the unit cell is preferred to be doubled along the *a*-axis in the case of tensile strain larger than 1%.

According to the definition, $TbMnO_3$ experiences 1% compressive strain when grown on the cubic substrate STO. Our calculations show that its ground state is ferromagnetic (see Fig. 1), in agreement with experiments [13]. This FM ground state is also consistent with the symmetric exchange interactions between spins of $Mn^{3+}$ ions, which are defined in Ref. 11. Our calculated results [21] show that the in-plane nearest neighboring interaction $J_{ab}$ is strongly FM and its magnitude is much larger than that of the in-plane next nearest neighboring interaction $J_{aa}$ and $J_{bb}$, indicating that the spins of $Mn^{3+}$ ions align ferromagnetically in the (001) planes. Interestingly, the magnetic interaction along *c*-direction is also FM, different from the bulk case.

The appearance of ferromagnetism in epitaxially strained $TbMnO_3$ is due to the change of the electronic structure as discussed below. In bulk $RMnO_3$, each $Mn^{3+}$ ion has four unpaired *d* electrons: three of them occupy the low-energy $t_{2g}$ orbitals and the remaining one occupies the high-energy $e_g$ orbitals. Because of the Jahn-Teller distortion, the two degenerate $e_g$ states on each spin site split into two non-degenerate orbitals and only one is occupied. The occupied $e_g$ orbitals on two adjacent spin sites are almost orthogonal to each other in the *ab*-plane. However, the orbital order in epitaxially strained $TbMnO_3$ is revealed to be dramatically different:

the two $e_g$ orbitals on each spin site are almost degenerate and can be seen as both half occupied (see Fig. 2), i.e., the electron configuration of each $Mn^{3+}$ ($d^4$) ion can be roughly described as $(t_{2g})^3(d_{3z^2-r^2})^{\frac{1}{2}}(d_{x^2-y^2})^{\frac{1}{2}}$. This electron configuration is consistent with the shape of the $MnO_6$ octahedron (see the insert in Fig. 2): All the Mn-O bond lengths are of similar magnitude and the O-Mn-O bond angles are close to $90^o$ or $180^o$. Now, let us examine the interaction between the $e_g$ orbitals of two adjacent spin sites (*1* and *2*) along *c*-direction for the cases of both FM and AFM arrangements. In the FM case (see Fig. 2), the majority-spin $e_g$ ($d_{z^2}$) orbitals of site *1* and *2* with the same energy level can couple with each other, forming an empty higher energy level and an occupied lower energy level, thereby lowering the total energy by *t*, where *t* (>0) is the hopping integral between these two $d_{z^2}$ orbitals. In the AFM case (see Fig. 2), the majority-spin $e_g$ orbital of one Mn site couples the minority-spin $e_g$ orbital of the other Mn site. In this case, the total energy lowering is $\frac{U}{2}-\sqrt{t^2+\frac{U^2}{4}}$, where U (>0) is the energy difference between $e_g$ orbitals with opposite spin components. It is clear that the FM state has a lower energy. The same mechanism also applies to the in-plane exchange interaction $J_{ab}$. Thus, the epitaxially strained $TbMnO_3$ prefers to interact ferromagnetically. In addition, the nearest neighboring Mn-Mn distances decrease and the Mn-O-Mn angles become larger with respect to those of bulk, which enhance the coupling of $e_g$ orbitals (increase of *t*) and thereby increase the magnitude of spin exchange interaction parameters. This is consistent with the increasing of the symmetric spin exchange parameters [21].

We now discuss the ground states as the epitaxial strain varies. The dependence of the total energy of one formula unit (f.u.) upon the epitaxial strain is shown in Fig. 1. Our calculated results show that the C-AFM and G-AFM spin orders almost always have higher energy than others (not shown in Fig. 1. for simplicity). In the compressive epitaxial strain region, TbMnO$_3$ is metallic FM, distinctly different from the insulating AFM bulk. At the 0% strain, there is a first-order isosymmetric phase transition [22] between two *Pbnm*-FM states, which is also confirmed by analyzing the Jahn-Teller distortion (see Part 2 of [18]). A gap opens when the tensile strain is higher than 1% (see the inset of Fig. 1).

Interestingly, when the tensile strain is larger than $\cong$ 4.5%, the ground state is E-type ordering with a polar *Pmc2$_1$* space group. In this E-AFM order, the FM zigzag chain is along the *b*-axis (Fig. 3a), which is different from that along the *a*-axis in the *Pbnm*-E-AFM HoMnO$_3$ case [23]. The evolution of the geometrical structure with E-AFM spin order is shown in Figs. 3(d) and 3(e). Before the E-AFM spin order is adopted as the ground state, there is only a small difference between the in-plane Mn-Mn bond distance along zigzag chains and that between two neighboring zigzag chains. However, when TbMnO$_3$ adopts the E-AFM spin order as the ground state, the difference is as large as 0.8 Å. Moreover, the Mn-O-Mn bond angles $\alpha_p$ [see Figs. 3(a)] along zigzag chains are closer to $180^o$ than the Mn-O-Mn bond angles $\alpha_{ap}$ [see Fig. (3a)] between two neighboring zigzag chains, which is similar to the case of *Pbnm*-E-AFM HoMnO$_3$ [23]. This is due to the fact that the $180^o$ Mn-O-Mn bond leads to an increased hopping between the occupied $e_g$ orbital and the unoccupied $e_g$ orbital. It is worth noting that an O-Mn-O bond angle severely deviates from

$180^0$ in the MnO$_6$ octahedron of *Pmc2$_1$*-E-AFM TbMnO$_3$, different from the case of *Pbnm*-E-AFM HoMO$_3$ [see Figs. 3(b) and 3(c)]. These indicate that epitaxially strained *Pmc2$_1$* TbMnO$_3$ displays substantial lattice distortions.

In order to understand why this new *Pmc2$_1$*-E-AFM state becomes the ground state, we decompose the total energy $E_{tot}$ into the elastic energy $E_e$ (including Coulomb electrostatic interaction, i.e., Madelung energy), and the magnetic energy $E_m$ originating from spin exchange interactions, namely $E_{tot} = E_e + E_m$. In *Pmc2$_1$* structure, four different in-plane exchange paths (see Fig. 3a) and two different paths along *c*-direction (not shown) are considered, i.e., J$_{ab1}$, J$_{ab2}$, J$_{aa}$, J$_{bb}$, J$_{cc1}$ and J$_{cc2}$. By calculating spin exchange parameters and the total energy, we can obtain the elastic energy $E_e$ and magnetic energy $E_m$. The calculated results in the case of 5% tensile strain are listed in Table *I*. While the magnetic energy $E_m$ of *Pbnm*-FM state is lower than that of *Pmc2$_1$*-E-AFM state, the elastic energy $E_e$ of *Pbnm*-FM state is much higher than that of *Pmc2$_1$*-E-AFM state. This indicates that the *Pmc2$_1$* structure is much softer than the *Pbnm* structure, which is consistent with the results on tensile strained BiFeO$_3$ [20] and with that *Pbnm* structure has a higher Madelung energy than *Pmc2$_1$* structure (see Part 4 of [18]). In addition, *Pmc2$_1$*-FM state has a higher magnetic energy $E_m$ than that of *Pmc2$_1$*-E-AFM state, which indicates that in the *Pmc2$_1$* structure the E-AFM spin order is preferred over the FM spin order. Hence, this new *Pmc2$_1$*-E-AFM state found in the tensile strained TbMnO$_3$ is a cooperative result of the softness of the *Pmc2$_1$* structure and compatible spin exchange interactions.

This new *Pmc2₁*-E-AFM state is insulating and polar, thus it could be ferroelectric. Using the E-AFM spin order [23], our Berry phase calculation shows that the electric polarization is as large as 4.56 µC/cm$^2$ along the zigzag chain, i.e., the positive *b*-axis. The large electric polarization in *Pmc2₁*-E-AFM state is almost two orders of magnitude larger than that in bulk TbMnO$_3$. It was well-known that the total electric polarization ($P_{tot}$) in multiferroics contains both the electronic contribution ($P_{ele}$) and the ionic contribution ($P_i$), i.e. $P_{tot} = P_{ele} + P_i$. To separate these two contributions, we directly calculate the electronic contribution $P_{ele}$ due to the E-AFM spin order in this new *Pmc2₁*-E-AFM state using the model proposed in Ref. 2 and get that the electronic contribution $P_{ele}$ is $-0.64 \, \mu C/cm^2$. Thus, the ionic contribution $P_i$ due to ion displacements from centrosymmetric positions is $5.20 \, \mu C/cm^2$, dominating in the total electric polarization $P_{tot}$. Note that the non-negligible negative electronic contribution will result in a negative magnetoelectric effect [24]. This is dramatically different from the case of *Pbnm*-E-AFM HoMnO$_3$ where the electronic contribution is in the same direction and has a similar magnitude as the ionic contribution [23]. Detailed calculations (see Part 3 of [18]) reveal that the positive ionic contribution is dominated by Terbium ions' contributions. Further investigations evidence that such positive ionic contribution is favored by Coulomb electrostatic interaction (see Part 4 of [18]). As for the negative electronic contribution, it is due to the homogenous positive *b*-direction migration of $e_g$ electrons from the occupied $d_{z^2}$ orbital of one $Mn^{3+}$ ion to the unoccupied $d_{x^2-y^2}$ orbital of its nearest neighboring $Mn^{3+}$ ion in the FM zigzag chains (see Part

5 of [18]).

We now examine the magnetic transition temperature. The Neel temperature $T_N$ of this new $Pmc2_1$-E-AFM state can be approximated by solving $E_H$ with mean field theory [25]: $T_N = \frac{2}{3}S(S+1)(J_1z_1 - J_2z_2 - J_3z_3 - J_4z_4 + J_5z_5 - J_6z_6) \approx 247$ K. We also perform Monte Carlo (MC) simulations to find that the transition temperature $T_N$ is about *100 K* (see Part 6 of [18]). It is well-known that the mean field theory overestimates and the MC simulation underestimates the transition temperature, respectively. Thus, we expect that the Neel temperature is close to 150 K. Compared with bulk TbMnO$_3$ ($T_N \approx 28\ K$), the transition temperature in this new $Pmc2_1$-E-AFM state is much higher.

Finally, there are two different E-type spin orders (E1 and E2) [23, 26], i.e. two different AFM domains, in the tensile strained TbMnO$_3$. These two AFM domains have equal-magnitude but opposite electric polarizations. In addition, these two domains have one-to-one correspondence relationship with the directions of electric polarizations, denoted as $+\mathbf{P}(\mathbf{E1})$ and $-\mathbf{P}(\mathbf{E2})$ (see Part 7 of [18]). We investigated the switching between them. In HoMnO$_3$, a $180^0$ coherent progressive rotation of spins of Mn$^{3+}$ ions was proposed to switch these two domains [26]. However, this mechanism may be not suitable to the case of the epitaxially strained TbMnO$_3$ because the polarization is dominated by ion displacements but not by the E-AFM order. Hence, we propose another suitable mechanism in the structural space. We starts with $+\mathbf{P}(\mathbf{E1})$ lattice structure, then the ions' positions are slightly displaced step by step toward to $-\mathbf{P}(\mathbf{E2})$ lattice structure. This path must pass through a centrosymmetric paraelectric phase. In each movement step, the magnetic order with the lowest energy for a giving geometrical structure is considered. Using this strategy, we find that the depth of the well is ≈ 75 meV/ f.u. (see Part 7 of [18]), which is smaller than the value ≈ 200 meV/f.u. in the case of $PbTiO_3$ [27]. In fact, the calculated energy barrier is an upper limit for experiments. Hence, we show that it's possible to use an external electric field to switch the two E-AFM domains experimentally.

In summary, our first-principles results show that, although unstrained bulk $TbMnO_3$ prefers to be insulating antiferromagnetic, compressed strained $TbMnO_3$ can be ferromagnetic, in agree with experiments [13]. The appearance of ferromagnetism is attributed to the change of electronic structure with respect to bulk $TbMnO_3$. Of the most importance is that, when $TbMnO_3$ experiences an epitaxial strain larger than ≈ 4.5%, a phase transition from paraelectric ferromagnetic order into ferroelectric E-type antiferromagnetic order takes place, resulting in a polarization as large as 4.56 $\mu C/cm^2$ and a relatively high transition temperature above *100 K*. Furthermore, we found that in the tensile strained $TbMnO_3$, the E-type magnetic order is coupled with the polar order, thus it is possible to manipulate the two different antiferromagnetic domains through an external electric field. These show that one can engineer $TbMnO_3$ as fascinating perovskite multiferroic material by substrate-induced epitaxial strain.

Work at Fudan was partially supported by NSFC, the Special Funds for Major State Basic Research, Foundation for the Author of National Excellent Doctoral

Dissertation of China, The Program for Professor of Special Appointment at Shanghai Institutions of Higher Learning, Research Program of Shanghai municipality and MOE. We thank X. Z. Lu and Z. L. Li for useful discussions.

E-mail: xggong@fudan.edu.cn, hxiang@fudan.edu.cn

(Saunders Company, Philadelphia, 1966).

**FIGURES**

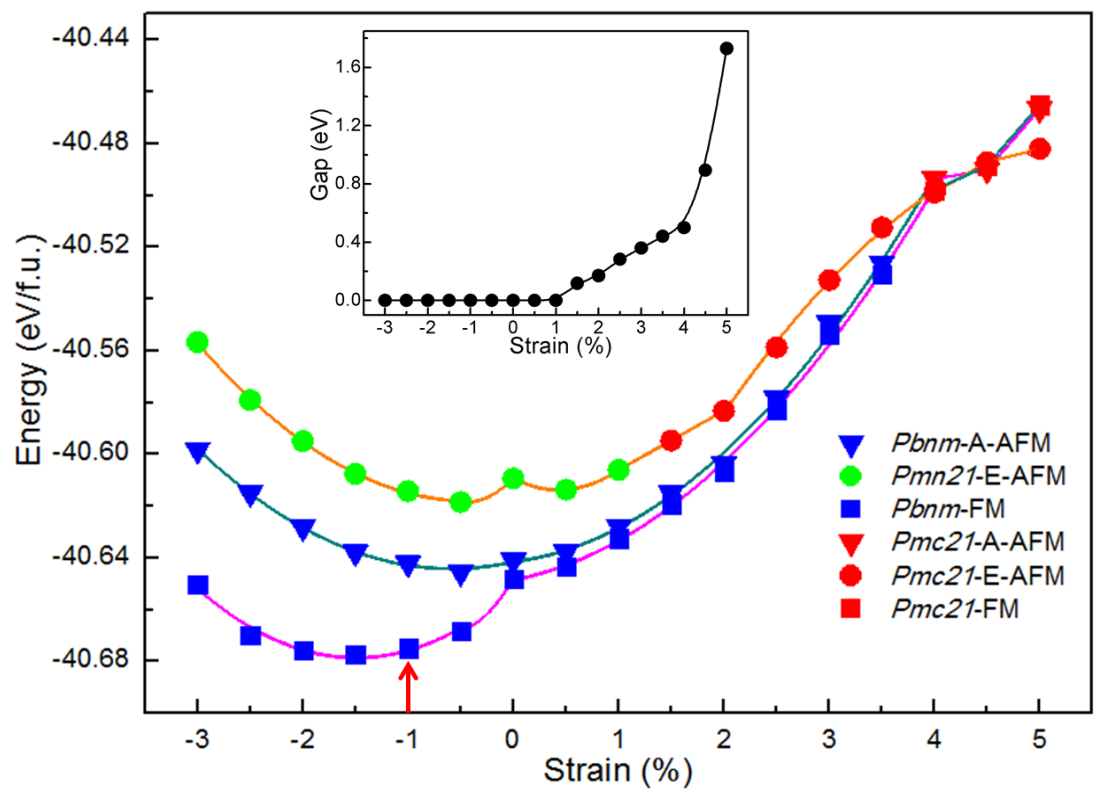

FIG. 1 (color online). Calculated total energy and band gap (insert) versus strain. Circle, down triangle and square represent E-AFM, A-AFM, FM spin orders, respectively. Red, green and blue colors label *Pmc2$_1$*, *Pmn2$_1$* and *Pbnm* space group respectively. Red arrow hightlights the situation where (001)-oriented TbMnO$_3$ is epitaxially grown on the (001) plane of cubic substrate SrTiO$_3$.

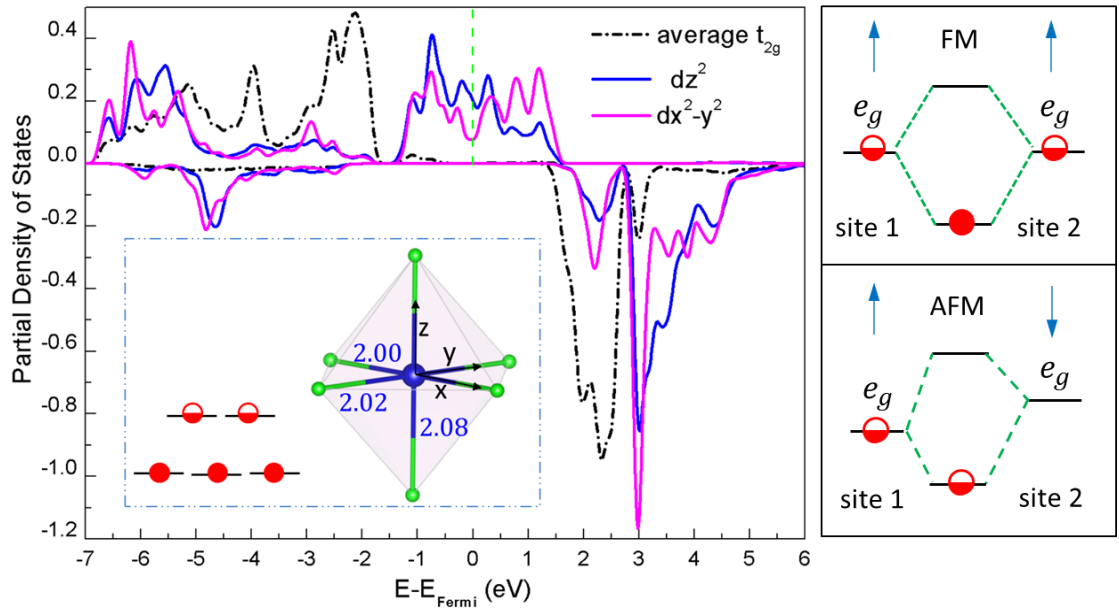

FIG. 2 (color on line). Left panel shows the Mn $d$-orbital partial density of state (PDOS). For simplicity, we plot the average $t_{2g}$ PDOS. Insert are $MnO_6$ octahedron and electron configuration. Numbers give the Mn-O bond length in Å. Green vertical line gives Fermi energy level. Righ panel illustrates schematically the hybridization between $e_g$ orbitals in FM (top) and AFM (bottom) cases. Half-filled circle represents half an electron. Spins are represented by arrows.

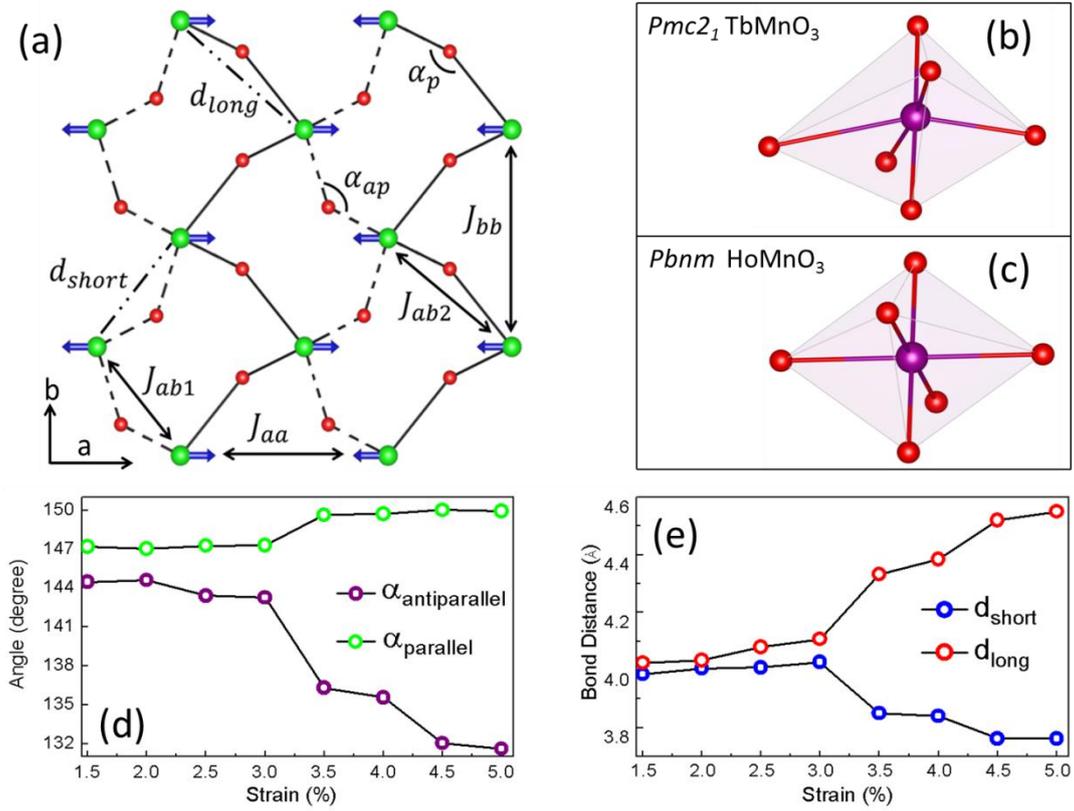

FIG. 3 (color online). (a) Illustration of E-type zigzag chains (solid lines). Large violet and small red spheres represent Mn and oxygen respectively. $\alpha_p$, $\alpha_{ap}$ bond angles (see text), short ($d_{short}$), long ($d_{long}$) Mn-Mn bonds, and exchange paths are denoted. MnO$_6$ octahedra of $Pmc2_1$ TbMnO$_3$ and bulk $Pbnm$ HoMnO$_3$ are shown in (b) and (c), respectively. (d) Evolution of $\alpha_a$, $\alpha_{ap}$ versus strain. (e) Evolution of $d_{short}$ and $d_{long}$ versus strain.

TABLE I. Total energy $E_{tot}$, elastic energy $E_e$ and magnetic energy $E_m$ for different states at 5% strain are listed.

| strain | states | $E_{tot}$(eV/f.u.) | $E_e$(eV/f.u.) | $E_m$(eV/f.u.) |
|---|---|---|---|---|
| +5% | *Pbnm*-FM | -40.4412 | -40.4243 | -0.0169 |
| +5% | *Pmc2₁*-FM | -40.4706 | -40.4688 | -0.0019 |
| +5% | *Pmc2₁*-E-AFM | -40.4818 | -40.4684 | -0.0134 |

Supplementary Materials for

A New Multiferroic State with Large Electric Polarization in Tensile Strained TbMnO$_3$

Y. S. Hou, J. H. Yang, X. G. Gong[*], and H. J. Xiang[*]

[1] Key Laboratory of Computational Physical Sciences (Ministry of Education), State Key Laboratory of Surface Physics, and Department of Physics, Fudan University, Shanghai 200433, P. R. China

1. Details of the density functional calculations

Our calculations were based on density functional theory (DFT) using generalized gradient approximation (GGA) plus the on-site Coulomb repulsion U method implemented in Vienna *ab initio* simulation package (VASP) [1-4]. In our calculation, an effective U value of 2 eV is applied to Mn *3d* states and a plane wave energy cutoff of 500 eV is used. Structural optimizations are performed toward equilibrium until the Hellmann-Feynman forces are less than 0.01 eV/Å. To calculate the electric polarization, we employ the Berry phase method [5].

2. Isosymmetric phase transition analysis

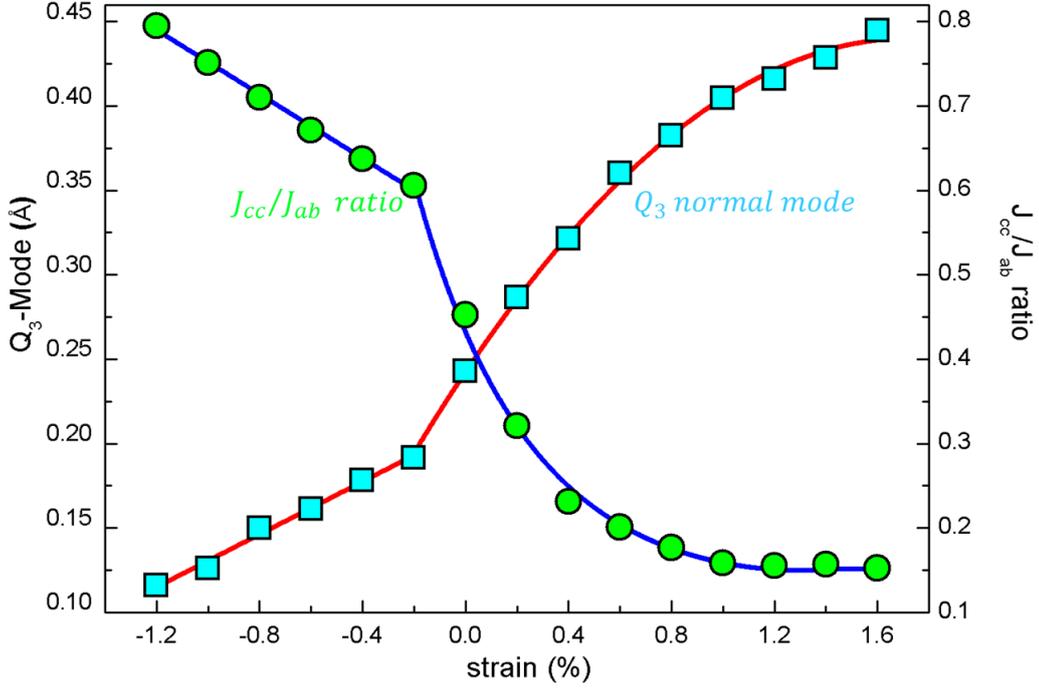

FIG. S1. $Q_3$ normal mode and $J_{cc}/J_{ab}$ magnetic coupling ratio versus strain are shown, where $J_{cc}$ is interlayer magnetic coupling and $J_{ab}$ is intralayer nearest neighboring magnetic coupling.

It is clear that, $Q_3$ normal mode [6] characterizing the Jahn-Teller distortion of $MnO_6$ octahedron, and magnetic coupling ratio, sharply jump at the strain close to zero. Hence, there exists a discontinuous change in the internal coordinates of epitaxially strained $TbMnO_3$, i.e. a large structural reconstruction associated with a substantial change in the relative magnetic coupling strength. Detailed geometric structure analysis shows that the $MnO_6$ octahedron of epitaxially compressed $TbMnO_3$ is more regular than that of epitaxially tensile $TbMnO_3$ in terms of Mn-O bonds and O-Mn-O bond angles. This suggests that the isosymmetric phase transition is related to the internal structural change [7].

3. The electric polarization induced by the ion displacements

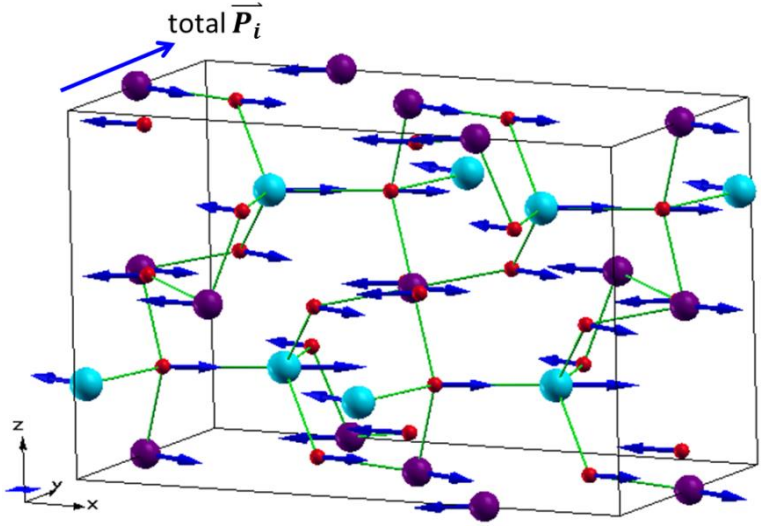

FIG. S2. Ion displacements and the total electric polarization $\vec{P}_i$ are shown. Blue arrow presents the direction and magnitude of an ion's displacement. Violet, blue and red spheres are Manganite, Terbium and Oxygen, respectively.

The electric polarization induced by ion displacements from the centrosymmetric positions can be estimated through the Born effective charges (BEC) method [8]. The estimated electric polarization induced by the Terbium, Manganese and Oxygen ions are found to be $10.22 \, \mu C/cm^2$, $-1.90 \, \mu C/cm^2$ and $-4.41 \, \mu C/cm^2$, respectively. Hence the estimated electric polarization induced by ion displacements is $3.91 \, \mu C/cm^2$, along positive *b*-direction, consistent with our DFT calculated results. The inconsistency of polarization magnitude between these two methods is attributed to the substantial structure distortion and thereby the linear approximation of BEC method is not very accurate.

4. Investigation of Coulomb electrostatic interaction

In order to understand why ion displacements induce the above-mentioned electric polarization, Coulomb electrostatic interactions, i.e., Madelung energies, of some different structure configurations, whose lattice constants are fixed to that of *Pmc2$_1$*-E-AFM structure in the case of 5% tensile strain, are calculated. Bader analysis indicates that Terbium, Manganese and Oxygen ions bear 2.21, 1.84 and -1.35 *e* respectively. Firstly, the centrosymmetric phase with *Pbnm* space group has a higher Madelung energy than that of *Pmc2$_1$*-E-AFM structure by 0.14 eV/ion. In addition, given that all the internal coordinates of Terbium ions in the *Pmc2$_1$*-E-AFM structure are replaced by that of centrosymmetric phase with *Pbnm* space group, denoting this structure as Tb-*Pbnm*-*Pmc2$_1$*, then the Tb-*Pbnm*-*Pmc2$_1$* structure has a higher Madelung energy than *Pmc2$_1$*-E-AFM structure by 0.07 eV/ion. Therefore, the ion displacements discussed above (and the electric polarization along positive *b*-direction) are favored by Coulomb electrostatic energy.

5. The electric polarization induced by the pure E-type magnetic order in HoMnO$_3$ and TbMnO$_3$, respectively

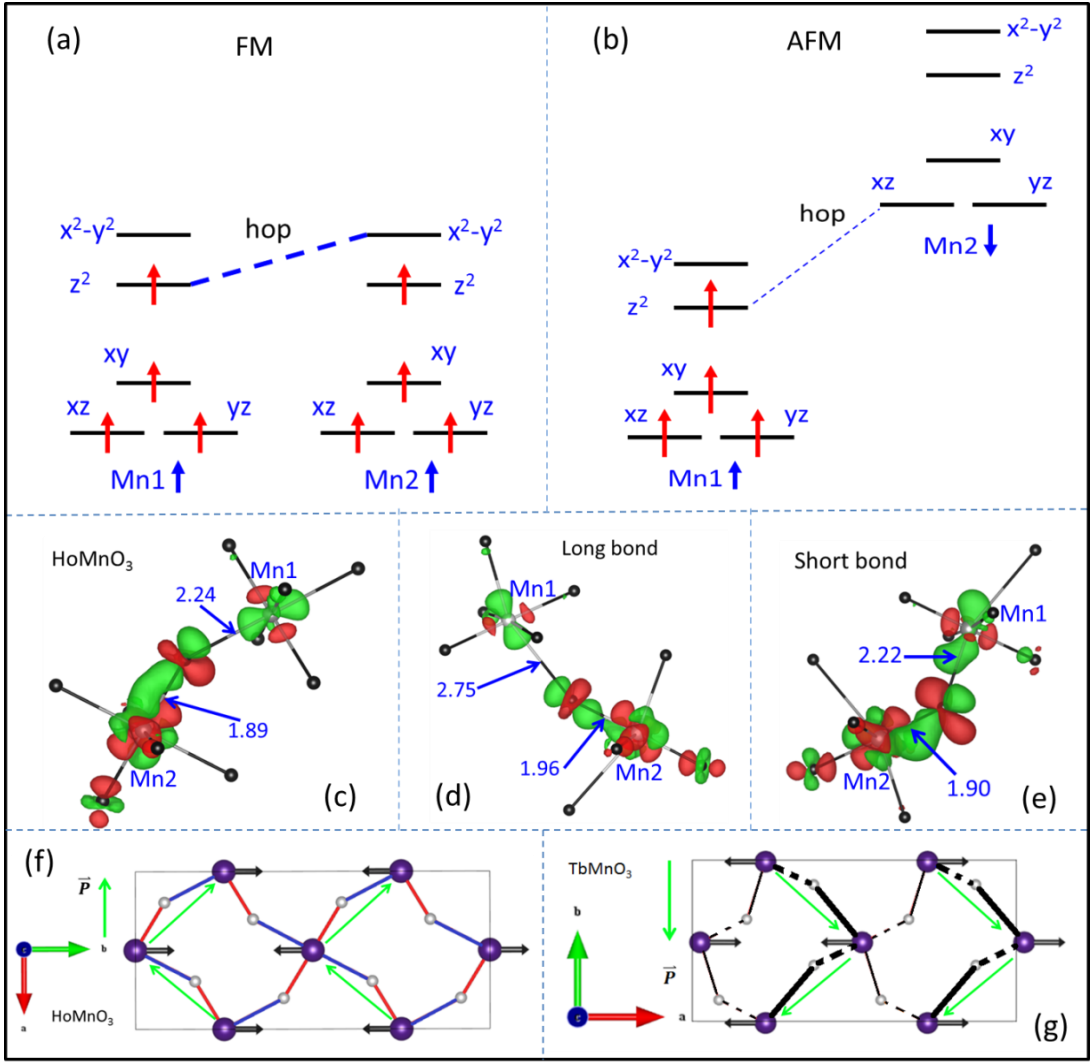

FIG. S3. Schematic illustration of the electron's hopping in the (a) FM case and (b) AFM case, respectively. Electron density difference $\Delta\rho = \rho(\uparrow,\uparrow) + \rho(\downarrow,\downarrow) - \rho(\downarrow,\uparrow) - \rho(\uparrow,\downarrow)$ for (c) bulk HoMnO$_3$, and (d) long Mn-Mn bond and (e) short Mn-Mn bond of the $Pmc2_1$ TbMnO$_3$ are plotted, where red color means $\Delta\rho > 0$ and the numbers give the Mn-O bond lengths in Å. Schematic illustrations of electric polarization induced by the pure E-type magnetic order are plotted for (f) bulk HoMnO$_3$ and (g) $Pmc2_1$ TbMnO$_3$, where green arrows depict the electric

polarization. In bulk HoMnO$_3$ (Fig. f), the red and blue lines stand for long and short Mn-O bonds, respectively. In *Pmc2$_1$*-E-AFM TbMnO$_3$ (Fig. g), solid lines and dashed lines stand for long and short Mn-O bonds, respectively.

First, let's investigate the *d*-electron hopping between two neighboring magnetic Mn$^{3+}$ ions. In FM case, the $e_g$ *d*-electron of Mn1 can hop to Mn2 (see Fig S3 a) because their *d*-orbital can strongly couple via the middle oxygen. However, in AFM case this hopping is negligible due to the very weak coupling since the coupled *d*-orbitals have a large energy difference (see Fig S3 b). This is also evidenced by the electron density difference plot shown in Figs S3 c, d and e. It is of importance to notice that, such hopping always leads to the homogenous migration of $e_g$ electrons from the occupied $d_{z^2}$ orbital of long Mn-O bond to the unoccupied $d_{x^2-y^2}$ orbital of short Mn-O bond. Based on this, pure E-type spin order produces an electric polarization along the negative *a*-direction in bulk HoMnO$_3$, consistent with the result of Ref 9, and along the negative *b*-direction in TbMnO$_3$ (see Figs S3 f and g), consistent with our DFT calculation results.

6. Monte Carlo simulation for the transition temperature in the case of 5% epitaxially strained TbMnO$_3$

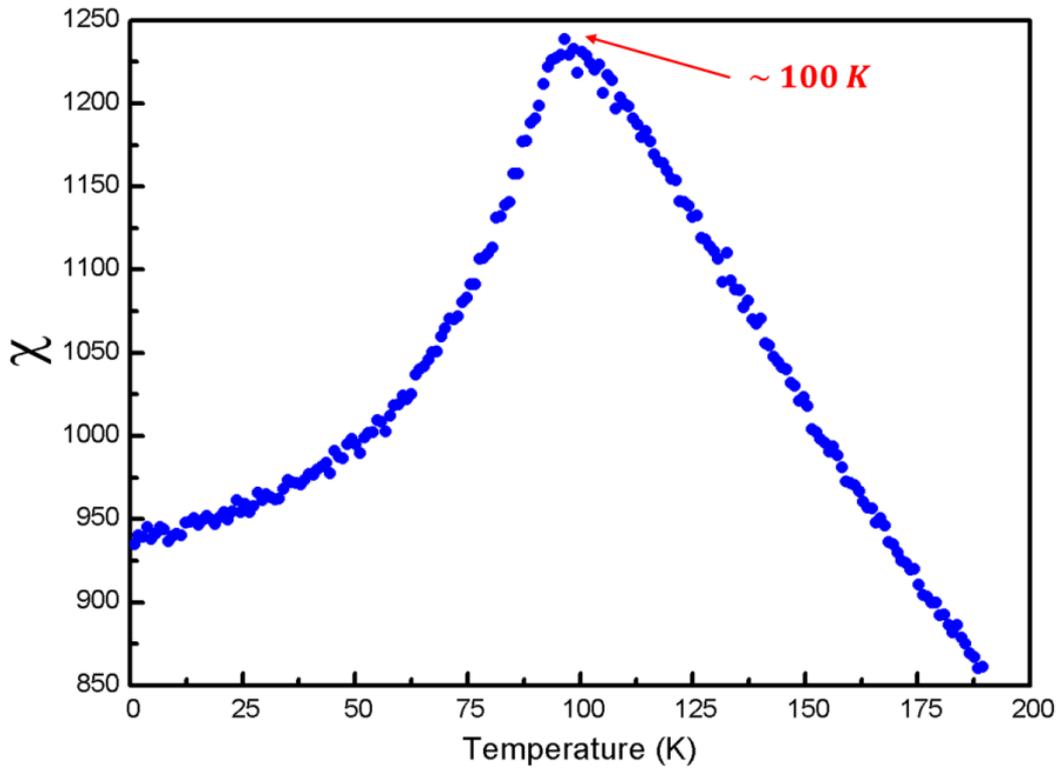

FIG. S5. Magnetic susceptibility as a function of temperature from the Monte Carlo simulation. Clearly, a peak appears at about *100 K*, indicating a phase transition from the paramagnetic state to the E-AFM state. It is well-known that the classical Monte Carlo underestimates the transition temperature. Thus, we expect that the transition temperature of the *Pmc2$_1$*-E-AFM state is above 100 K.

7. Double well structure for *Pmc2₁*-E-AFM state in the case of 5% epitaxially strained TbMnO$_3$

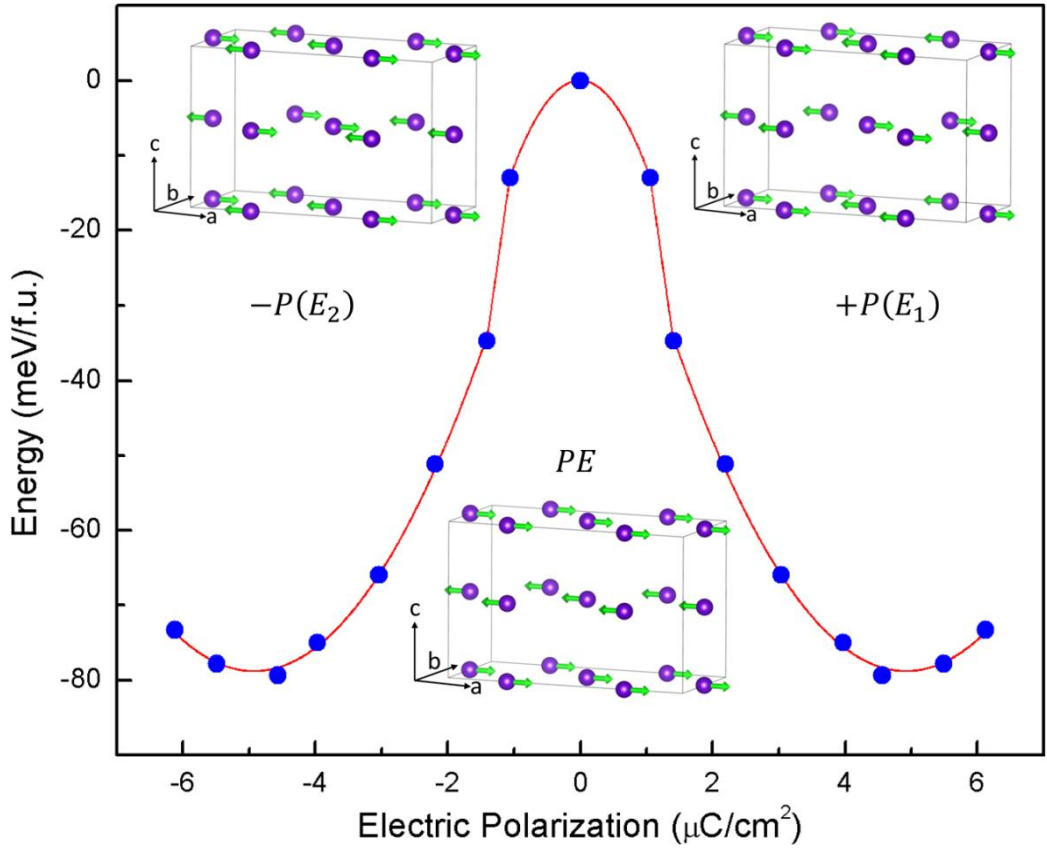

FIG. S6 (color online). Calculated electric polarization ($\mu$C/cm$^2$) versus energy (meV/formula unit). Inserts are antiferromagnetic domains $-\mathbf{P}(\mathbf{E}2)$, $+\mathbf{P}(\mathbf{E}1)$ and antiferromagnetic paraelectric phase respectively. Only Mn$^{3+}$ ions are shown.